\documentclass[aps,prl,floatfix,twocolumn,showpacs,superscriptaddress,groupedaddress]{revtex4-2}
\usepackage{graphicx}%
\usepackage{amsmath, amssymb}
\usepackage{color}
\usepackage{amsfonts}%
\usepackage{hhline}
\usepackage{braket}
\usepackage{bm}
\usepackage{mathrsfs}
\usepackage[normalem]{ulem}
\usepackage{hyperref}
\hypersetup{colorlinks=true,breaklinks,linkcolor=blue,urlcolor=blue,citecolor=blue}

\newcommand{\green}[1]{\textcolor{red}{#1}}
\definecolor{brown}{rgb}{0.55,0.30,0.05}

\def\vec#1{\boldsymbol #1}
\newcommand{\Ha}{\mathcal{H}}
\newcommand{\PG}{\mathcal{P}_{\mathrm{G}}}

\newcommand{\PJ}{\mathcal{P}_{\rm J}}
\newcommand{\Ns}{N_{\rm s}}
\newcommand{\Ne}{N_{\rm e}}

\definecolor{tmdgreen}{rgb}{0.0,0.42,0.16}

\begin{document}
\title{
Exotic superconductivity in the doped Kitaev quantum spin liquid
}
\author{Takahiro Misawa$^1$}
\author{Kota Ido$^2$}
\affiliation{
$^1$Institute for Solid State Physics, University of Tokyo, Kashiwa, Chiba 277-8581, Japan\\
$^2$Department of Applied Science, Graduate School of Sciences and Technology for Innovation, Yamaguchi University,
2-16-1 Tokiwadai, Ube, Yamaguchi 755-8611, Japan 
}

\begin{abstract}
We investigate superconductivity in a doped Kitaev quantum spin liquid by applying the many-variable variational Monte Carlo method to the hole-doped $t$-$J$-type Kitaev model.
Using a projected pair-product wave function that can exactly represent the Kitaev quantum spin liquid, we examine the stability of superconducting phases on isotropic two-dimensional clusters. For the ferromagnetic Kitaev interaction, robust triplet $p$-wave superconductivity coexists with ferromagnetism in the low-to-intermediate doping regime but is suppressed as the system approaches the fully polarized ferromagnetic phase.
For the antiferromagnetic Kitaev interaction, superconductivity exhibits a change in the dominant pairing symmetry from spin-dependent triplet $p$-wave at low doping to singlet $d+id$ at intermediate doping.
By varying the strength of the ferromagnetic Kitaev interaction at fixed doping, we show that the triplet superconductivity increases together with the ferromagnetic moment and becomes strongest slightly below full polarization.
Our results provide a theoretical basis for experimental searches for unconventional superconductivity, such as triplet superconductivity coexisting with ferromagnetism, in carrier-doped Kitaev candidate materials.
\end{abstract}

\maketitle

{\it Introduction.---}%
\label{sec:introduction}
The Kitaev model on the honeycomb lattice provides a rare example of an exactly solvable two-dimensional quantum spin model whose ground state is a quantum spin liquid with Majorana fermions as low-energy degrees of freedom~\cite{Kitaev_ANP2006,Baskaran_PRL2007}.
Emergent Majorana fermions in quantum magnets have attracted considerable interest as building blocks for topological quantum computation~\cite{Kitaev_AP2003,Freedman_BA2003}, and intensive efforts have been devoted to realizing the Kitaev quantum spin liquid in materials.
Since the Kitaev interaction was predicted to become dominant in Mott insulators with strong spin-orbit coupling and edge-sharing octahedral geometry~\cite{Jackeli_PRL2009},
the search for materials guided by this principle has been actively pursued~\cite{Winter_2017rev,TakagiTJKN2019,TrebstH2022,Matsuda_RMP2025,Motome_JPSJ2020}.
In particular, reports of half-quantized thermal Hall conductivity in $\alpha$-RuCl$_3$ have been regarded as strong evidence for Majorana fermions~\cite{Kasahara_Nature2018,Yokoi_Science2021}, although the interpretation remains under active debate~\cite{Yamashita_PRB2020,Hentrich_PRB2019,Lefrancois_PRX2022,Czajka_NMat2023}.
This observation has nevertheless further stimulated the search for Kitaev materials.

Beyond the undoped limit, carrier doping of quantum spin liquids has long been discussed as a route to unconventional superconductivity since the resonating-valence-bond (RVB) proposal for cuprates~\cite{Anderson_Science1987,Lee_RMP2006}.
Because the Kitaev quantum spin liquid is an exact RVB state, namely a projected BCS state~\cite{Burnell_Nayak_PRB2011,Schaffer_2012, Fu_PRB2018, Udagawa_JPCM2021}, its carrier doping offers an ideal realization of this scenario.
In this context, recent experimental and \textit{ab initio} studies of van der Waals heterostructures, particularly $\alpha$-RuCl$_3$/graphene and $\alpha$-RuCl$_3$/graphite interfaces, have provided evidence for sizable interfacial charge transfer and proximity-induced reconstruction of the electronic states, suggesting a realistic route toward carrier-doped Kitaev candidate materials~\cite{Mashhadi_NL2019,Zhou_PRB2019,Wang_NL2020,Rizzo_NL2020,Rossi_NL2023,Zheng_PRB2023,Zheng_NC2024,Biswas_PRL2019,Gerber_PRL2020}.

Previous theoretical studies of doped Kitaev systems have explored the behavior of mobile carriers~\cite{Mei_PRL2012,Halasz_PRB2014} and predicted unconventional superconducting states, including triplet $p$-wave and singlet $d+id$ states~\cite{You_PRB2012,Hyart_PRB2012,Okamoto_PRB2013,Scherer_PRB2014}.
However, studies beyond mean-field theory remain limited.
Recent density-matrix renormalization group (DMRG) studies on quasi-one-dimensional geometries have revealed various magnetic, charge-ordered, and pairing tendencies in doped Kitaev systems~\cite{Kadow_npjQM2024,Peng_npjQM2021,Jin_npjQM2024,Laurell_PRB2024,Sousa_arXiv2025,Agrapidis_PRB2026,Pandey_arXiv2026}.
These studies have emphasized the importance of competing magnetic and charge correlations generated by carrier motion, although the resulting pairing tendencies depend sensitively on the lattice geometry and the kinetic energy scale.
Therefore, it is still challenging to examine the stability of superconducting as well as magnetic phases on isotropic two-dimensional clusters.%

\begin{figure}[t]
  \includegraphics[width=\columnwidth]{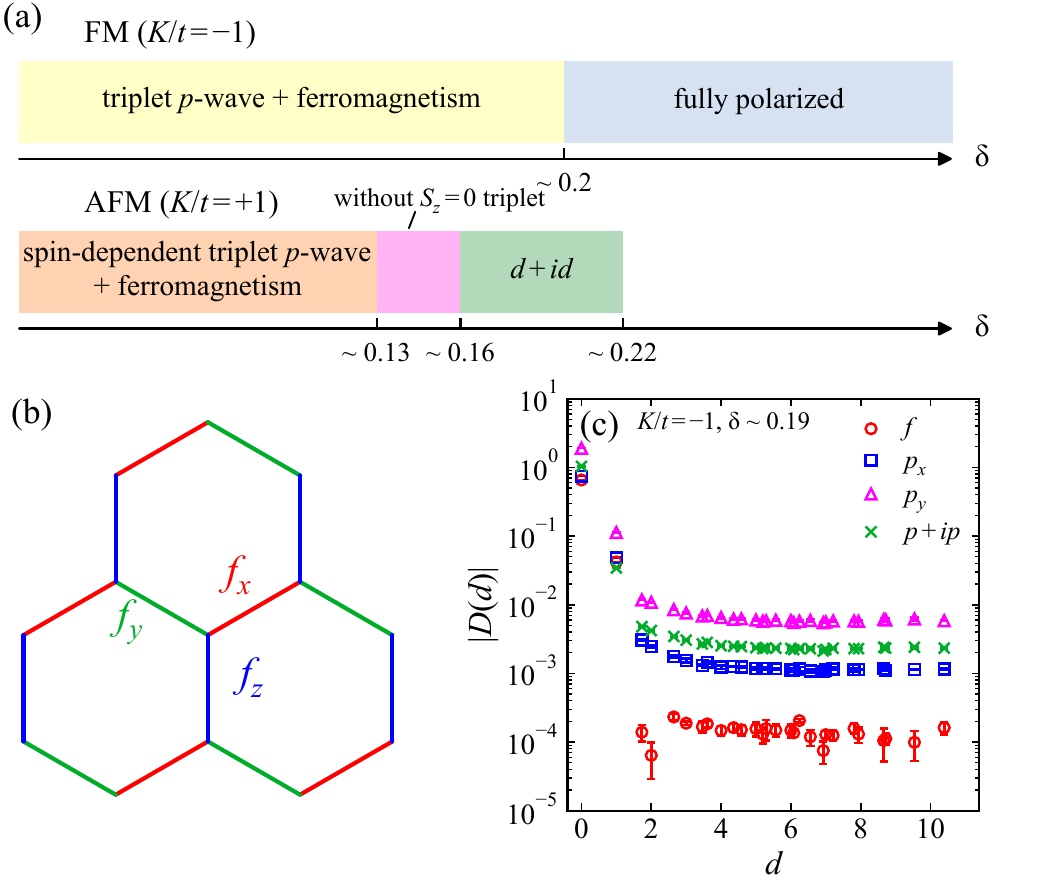}
  \caption{%
    (a)~Schematic phase diagrams as functions of the hole concentration $\delta$ for the ferromagnetic ($K/t=-1$) and antiferromagnetic ($K/t=+1$) Kitaev interactions. The shaded regions use the same colors as Figs.~\ref{fig:sc_mall_ferro} and \ref{fig:sc_mall_af}, and the phase boundaries are estimated from the $L=12$ results.
    (b)~Honeycomb lattice with the three bond types $\alpha\in\{x,y,z\}$ and the corresponding nearest-neighbor form factors $f_x$, $f_y$, $f_z$ used to construct the pairing operators.
    (c)~Distance dependence of the pairing correlation $D^\eta_\alpha(\vec{r})$ for the four triplet form factors ($f$, $p_x$, $p_y$, $p+ip$) in the $S_z=0$ sector on the $L=12$ cluster at $\delta\simeq 0.19$ with the ferromagnetic Kitaev interaction ($K/t=-1$).
    The plateau for $d \geq L/2$
    defines the long-range average $P^\eta_\alpha$.
  }
  \label{fig:model_pr}
\end{figure}

In this Letter, we study the hole-doped $t$-$J$-type Kitaev model on isotropic two-dimensional clusters with $L=8$, $10$, and $12$ using the many-variable variational Monte Carlo (mVMC) method with a projected pair-product (Pfaffian) wave function~\cite{Tahara_Imada_JPSJ2008,Misawa_CPC2019,Becca_Sorella_Book2017}, which can exactly represent the Kitaev quantum spin-liquid state~\cite{Burnell_Nayak_PRB2011,Schaffer_2012,Fu_PRB2018,Udagawa_JPCM2021} (see Supplemental Material Sec.~\ref{app:kqsl_pfaffian}~\cite{SupplementalMaterial}). Although this model is a simplified description of doped candidate materials, it captures the essential physics of carrier doping in the Kitaev quantum spin liquid. For the ferromagnetic interaction, we find triplet $p$-wave superconductivity that coexists with ferromagnetism before full polarization, whereas the antiferromagnetic interaction induces a transition in the dominant pairing symmetry from spin-dependent triplet $p$-wave to singlet $d+id$ pairing (see Fig.~\ref{fig:model_pr}(a)).
We further show that, at fixed low doping in the ferromagnetic case, the triplet correlation depends nonmonotonically on the Kitaev coupling, whereas the magnetization decreases monotonically with increasing coupling strength.
The coexistence of spin-triplet superconductivity and ferromagnetism found here offers a doped-spin-liquid counterpart of the ferromagnetic superconductors discussed in $f$-electron systems~\cite{Aoki_JPSJ2019}.

{\it Model and method.---}%
\label{sec:model_method}
We consider the hole-doped $t$-$J$-type Kitaev model on the honeycomb lattice,
\begin{align}
  \Ha
  &=
  -t\sum_{\langle i,j\rangle,\sigma}
  \left(
    c^{\dagger}_{i\sigma}c_{j\sigma} + {\rm H.c.}
  \right)
  + \sum_{\langle i,j\rangle_{\alpha}} K_{\alpha} S_i^{\alpha} S_j^{\alpha}.
  \label{eq:tJ_Kitaev}
\end{align}
Here $c^{\dagger}_{i\sigma}$ ($c_{i\sigma}$) creates (annihilates) an electron with spin $\sigma=\uparrow,\downarrow$ at site $i$, and $\langle i,j\rangle_{\alpha}$ denotes a nearest-neighbor bond of type $\alpha\in\{x,y,z\}$.
The spin operators are written in terms of the electron operators as
$S^{\alpha}_{i}=\frac{1}{2}\sum_{\sigma\sigma'}c^{\dagger}_{i\sigma}\tau^{\alpha}_{\sigma\sigma'}c_{i\sigma'}$,
where $\tau^{\alpha}$ are the Pauli matrices.
Details are shown in Supplemental Material Sec.~\ref{app:kqsl_pfaffian}.
Double occupancy in the $t$-$J$ model is excluded by the variational wave function introduced below [Eq.~\eqref{eq:variational_state}].
We consider the isotropic case $K_x=K_y=K_z= K$ and measure energies in units of $|t|$.
We define the hole concentration as $\delta = 1-\Ne/\Ns$, where $\Ne$ is the number of electrons and $\Ns$ the number of lattice sites.
We consider both the ferromagnetic ($K<0$) and antiferromagnetic ($K>0$) Kitaev interactions.

We perform calculations on honeycomb clusters with $\Ns=2L_xL_y$ sites (two sites per unit cell).
In the main results we focus on $L_x=L_y=L$ with $L=8, 10, 12$ ($\Ns=128, 200, 288$).
We employ antiperiodic--periodic (AP--P) boundary conditions for fermions.

Our variational wave function is defined as
\begin{align}
  \ket{\Psi}
  =
  \PJ \, \PG^{\infty} \ket{\phi_{\rm pair}},
  \label{eq:variational_state}
\end{align}
where $\ket{\phi_{\rm pair}}$ is the pair-product state~\cite{Kurita_PRB2015} given by
\begin{align}
  \ket{\phi_{\rm pair}}
  =
  \left(\sum_{I,J} F_{IJ}\, c_I^{\dagger} c_J^{\dagger}\right)^{\Ne/2}\ket{0},
  \label{eq:phi_pair}
\end{align}
with a skew-symmetric matrix $F_{IJ}=-F_{JI}$ and a combined index $I=(i,\sigma)$.
Here $\PG^{\infty}$ is the Gutzwiller projector~\cite{Gutzwiller_PRL1963} that fully prohibits double occupancy, enforcing the $t$-$J$ constraint.
The density--density Jastrow factor~\cite{Jastrow_PR1955,Capello_PRL2005} $\PJ$ is defined as $\PJ=\exp[-\frac{1}{2}\sum_{i\neq j} v_{ij} n_i n_j]$, where $n_i= \sum_\sigma c^\dagger_{i\sigma} c_{i\sigma}$.
We impose translational invariance on both $F_{IJ}$ and $v_{ij}$ under unit-cell translations.
All the variational parameters ($F_{IJ}$ and $v_{ij}$) are simultaneously optimized using the stochastic reconfiguration method~\cite{Sorella_PRB2001,Becca_Sorella_Book2017} implemented in the mVMC package~\cite{Tahara_Imada_JPSJ2008,Misawa_CPC2019}.
The initial conditions for the optimization and the procedure for selecting among competing optimized states are summarized in Supplemental Material Sec.~\ref{app:initial_states}.
At half filling, the pair-product wave function with the Gutzwiller projection can exactly represent the Kitaev quantum spin-liquid ground state~\cite{Burnell_Nayak_PRB2011,Schaffer_2012,Fu_PRB2018,Udagawa_JPCM2021}, as we detail in Supplemental Material Sec.~\ref{app:kqsl_pfaffian}.
There, we show that the variational energy at half filling agrees with the Majorana and exact-diagonalization results to within $10^{-5}$ per site for all clusters up to $L=20$.

To investigate superconductivity, we evaluate equal-time pairing correlations.
In the $S_z=0$ sector, we define the singlet ($S$) and triplet ($T$) pairing operators as
\begin{align}
  \Delta^{S/T}_{\alpha}(\vec{r}_i)
  =
  \frac{1}{\sqrt{2}}\sum_{\gamma} f^{\alpha}_{\gamma}
  \left(
    c_{\vec{r}_i\uparrow}\, c_{\vec{r}_i+\vec{e}_\gamma\downarrow}
    \mp c_{\vec{r}_i\downarrow}\, c_{\vec{r}_i+\vec{e}_\gamma\uparrow}
  \right),
  \label{eq:pair_op_ST}
\end{align}
where $\gamma$ runs over the three nearest-neighbor bond types $\{x, y, z\}$, $\vec{e}_\gamma$ is the nearest-neighbor vector along the $\gamma$ bond, and $(f^{\alpha}_{x},f^{\alpha}_{y},f^{\alpha}_{z})$ is the form factor for channel $\alpha$ [Fig.~\ref{fig:model_pr}(b)].
In the $S_z=\pm 1$ sector, we define equal-spin triplet pairing as
\begin{align}
  \Delta^{\sigma}_{\alpha}(\vec{r}_i)
  =
  \sum_{\gamma} f^{\alpha}_{\gamma}\,
    c_{\vec{r}_i\sigma}\, c_{\vec{r}_i+\vec{e}_\gamma\sigma}
  \quad (\sigma=\uparrow,\downarrow).
  \label{eq:pair_op_tt}
\end{align}
The pairing correlation is defined as
\begin{align}
  D^{\eta}_{\alpha}(\vec{r})
  =
  \frac{1}{\Ns}\sum_i
  \left\langle
    \Delta^{\eta\,\dagger}_{\alpha}(\vec{r}_i)\Delta^{\eta}_{\alpha}(\vec{r}_i+\vec{r})
  \right\rangle,
  \label{eq:pair_corr}
\end{align}
with $\eta\in\{S, T, \uparrow, \downarrow\}$.
Following the notation for nearest-neighbor pairing channels on the honeycomb lattice~\cite{Xu_PRB2016}, we use four form factors, each of which labels both a singlet and a triplet channel:
$(1,1,1)$ for $s$-wave (singlet) / $f$-wave (triplet),
$(1,-1,0)$ for $d_{xy}$ (singlet) / $p_x$ (triplet),
$(1,1,-2)$ for $d_{x^2-y^2}$ (singlet) / $p_y$ (triplet),
and $(\omega,\omega^{\ast},1)$ with $\omega=e^{2\pi i/3}$ for $d+id$ (singlet) / $p+ip$ (triplet).
Writing $\vec{r}=d_x\vec{a}_1+d_y\vec{a}_2$ with
$\vec{a}_1=(1,0)$ and $\vec{a}_2=(1/2,\sqrt{3}/2)$, we independently
reduce $d_x$ and $d_y$ to $(-L/2,L/2]$ and define
$d=|\vec{r}|=(d_x^2+d_xd_y+d_y^2)^{1/2}$.
For each distinct distance $d$, we define $D^\eta_\alpha(d)$ as
the value of $\operatorname{Re}D^\eta_\alpha(\vec{r})$ with the largest
absolute value over all $\vec{r}$ satisfying $|\vec{r}|=d$.
We define the long-range average of the pairing correlation as
\begin{align}
  P^{\eta}_{\alpha} = \frac{1}{N_{\rm dist}}
  \sum_{d \ge L/2}
  D^{\eta}_{\alpha}(d),
  \label{eq:pair_avg}
\end{align}
where the sum runs over distinct distances and $N_{\rm dist}$ is
the number of distances satisfying $d\ge L/2$.
To characterize magnetism, we evaluate the uniform magnetization per site,
\begin{align}
  M_{\alpha} = \frac{1}{\Ns}\sum_i \langle S^{\alpha}_i \rangle \quad (\alpha=x,y,z).
  \label{eq:mag_def}
\end{align}
We also define the total magnetization as $M_{\rm all}=(M_x^2+M_y^2+M_z^2)^{1/2}$.

\begin{figure}[t]
  \includegraphics[width=\columnwidth]{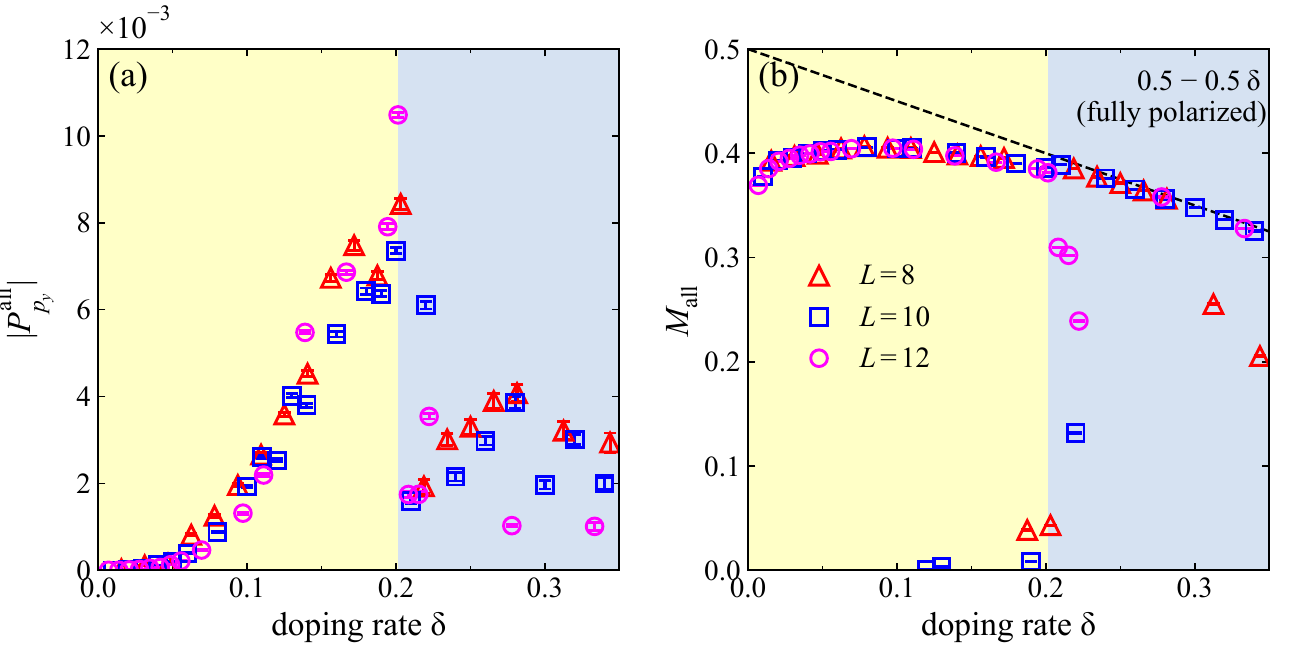}
  \caption{%
    Ferromagnetic Kitaev interaction ($K/t=-1$).
    (a)~Doping dependence of the long-range-averaged $p_y$ triplet superconducting correlation
    $P^{\rm all}_{p_y}=[(P^{T}_{p_y})^2+(P^{\uparrow}_{p_y})^2+(P^{\downarrow}_{p_y})^2]^{1/2}$
    for $L=8$, $10$, and $12$.
    (b)~Total magnetization $M_{\rm all}$ for $L=8$, $10$, and $12$.
    The dashed line shows the fully polarized value $(1-\delta)/2$.
    The yellow shading marks the doping range where the triplet $p$-wave superconducting correlation is enhanced.
    The blue shading marks the doping range where the magnetization approaches the fully polarized value.
  }
  \label{fig:sc_mall_ferro}
\end{figure}

\begin{figure}[t]
  \includegraphics[width=\columnwidth]{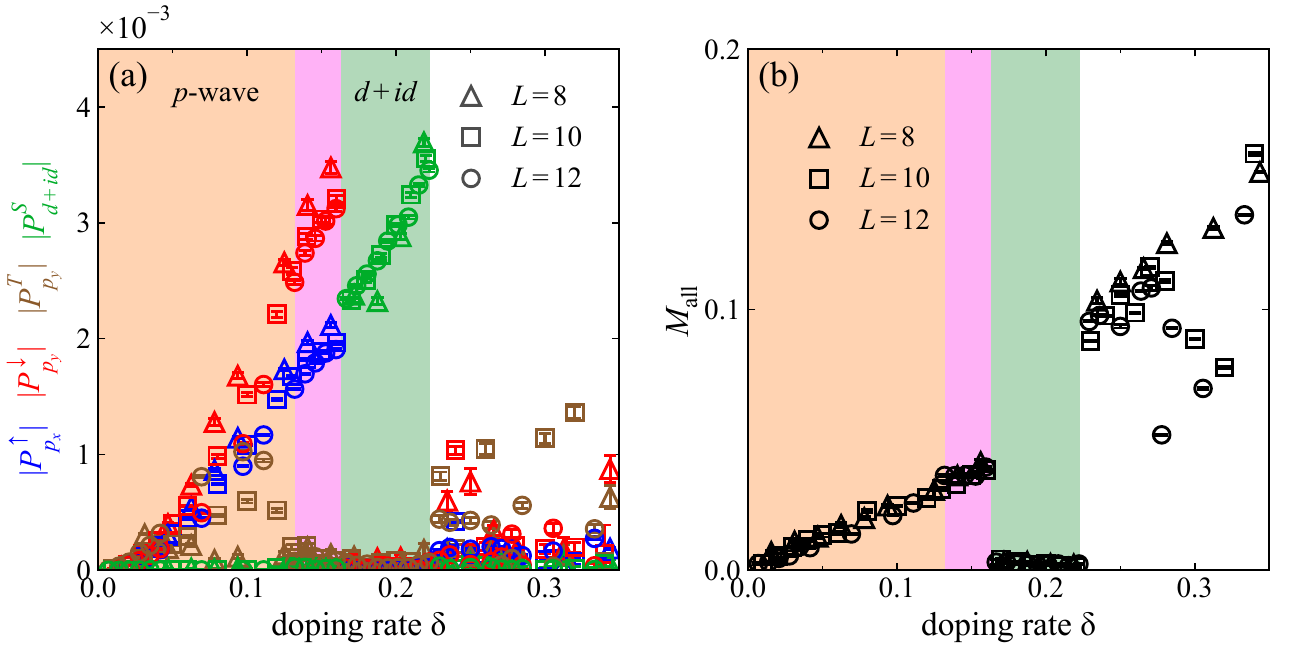}
  \caption{%
    Antiferromagnetic Kitaev interaction ($K/t=+1$).
    (a)~Doping dependence of representative long-range-averaged pairing correlations for $L=8$, $10$, and $12$:
    $S_z=0$ triplet $p_y$ (brown: $P^{T}_{p_y}$),
    equal-spin triplet (blue: $P^{\uparrow}_{p_x}$ and red: $P^{\downarrow}_{p_y}$),
    and singlet $d+id$ (green: $P^{S}_{d+id}$).
    Two triplet regimes appear at low doping, one with and one without sizable $P^{T}_{p_y}$, followed by a narrow $d+id$ regime and a high-doping regime with weak, size-dependent triplet correlations.
    The orange, magenta, and green shaded regions mark these two triplet regimes and the $d+id$ regime, respectively.
    (b)~Total magnetization $M_{\rm all}$ for $L=8$, $10$, and $12$.
  }
  \label{fig:sc_mall_af}
\end{figure}

{\it Results.---}%
\label{sec:results}
We first discuss the ferromagnetic Kitaev case ($K/t=-1$).
At half filling ($\delta=0$), the ground state is the Kitaev quantum spin liquid with no magnetic order.
Figure~\ref{fig:model_pr}(c) shows the distance dependence of the pairing correlation $D^\eta_\alpha(\vec{r})$ at $\delta\simeq 0.19$ on the $L=12$ cluster.
Since singlet correlations are small, we show the correlations for the four triplet form factors ($f$, $p_x$, $p_y$, $p+ip$).
The $p_x$- and $p_y$-wave components reach a plateau at large $d$ and show well-developed long-range superconducting correlations.
In the following analysis, we use the long-range average $P^\eta_\alpha$ [Eq.~\eqref{eq:pair_avg}] as a measure of the pairing strength.
Additional distance profiles supporting this long-range averaging procedure are shown in Supplemental Material Sec.~\ref{app:pairing_details}.
Throughout the figures, error bars indicate the statistical errors of the Monte Carlo sampling and are smaller than the symbol size for most data points.

Figure~\ref{fig:sc_mall_ferro}(a) shows the doping dependence of the $p_y$-wave triplet superconducting correlation,
$P^{\rm all}_{p_y}=[(P^{T}_{p_y})^2+(P^{\uparrow}_{p_y})^2+(P^{\downarrow}_{p_y})^2]^{1/2}$.
All three system sizes show the same behavior. The superconducting correlation increases rapidly with doping and reaches its maximum around $\delta\simeq 0.2$.
For $\delta\lesssim 0.2$, $P^{\rm all}_{p_y}$ shows only weak size dependence, indicating that triplet $p$-wave superconductivity is realized in this regime.
Beyond $\delta\simeq 0.2$, however, $P^{\rm all}_{p_y}$ is strongly suppressed and remains small up to the largest doping shown.
We note that the amplitudes of singlet correlations remain negligibly small in this doping range.

Figure~\ref{fig:sc_mall_ferro}(b) shows the doping dependence of the total magnetization $M_{\rm all}$.
Upon hole doping, ferromagnetism is induced even at very low hole concentration and persists over most of the plotted range.
For $\delta\gtrsim 0.2$, $M_{\rm all}$ closely follows the fully polarized value $(1-\delta)/2$ for all sizes.
In this saturated regime, both $M_x$ and $M_z$ become finite and the polarization axis tilts in the $x$--$z$ plane.
The sharp suppression of the triplet $p$-wave correlations occurs in the same doping range where the magnetization approaches this saturated value, indicating that the fully polarized state disfavors superconductivity.
The sizable triplet $p$-wave superconductivity is therefore confined to the low-to-intermediate doping regime before full polarization is reached.
In this regime, we find that triplet $p$-wave superconductivity coexists with ferromagnetism.
This extended ferromagnetic regime is consistent with recent DMRG results~\cite{Jin_npjQM2024}.
The narrow doping window of the triplet $p$-wave regime contrasts with the mean-field prediction of a broad stable superconducting phase~\cite{You_PRB2012,Hyart_PRB2012,Okamoto_PRB2013}, and is consistent with the view that full spin polarization suppresses the spin fluctuations needed to induce pairing.

A few points near $\delta\simeq 0.12$, $0.2$, and $0.3$ in Fig.~\ref{fig:sc_mall_ferro}(b) show strongly reduced magnetization. Optimization runs initialized with the $p{\rm SC}_2$ ansatz of Ref.~[\onlinecite{Okamoto_PRB2013}] [Supplemental Material Sec.~\ref{app:initial_states}] converge to nearly unpolarized superconducting states near $\delta\simeq 0.12$ and $0.2$, while partially depolarized states become energetically favorable at $\delta\simeq 0.2$ for $L=12$ and $\delta\simeq 0.3$ for $L=8$. Although this competition is strongly size dependent and its fate in the thermodynamic limit remains unclear, the dominant pairing channel remains triplet $p$-wave in all cases.

We next turn to the antiferromagnetic Kitaev case ($K/t=+1$).
Figure~\ref{fig:sc_mall_af}(a) shows the doping dependence of representative pairing correlations.
The data for the antiferromagnetic case can be classified into four regimes.
At the lowest doping levels, the triplet components $P^{T}_{p_y}$, $P^{\uparrow}_{p_x}$, and $P^{\downarrow}_{p_y}$ are selectively enhanced, while the remaining equal-spin components such as $P^{\uparrow}_{p_y}$ stay small.
The decomposition of all six $p_x$- and $p_y$-wave triplet components is shown in Supplemental Material Sec.~\ref{app:pairing_details}.
This spin- and form-factor-selective triplet pairing was not identified in previous mean-field phase diagrams~\cite{You_PRB2012,Hyart_PRB2012,Okamoto_PRB2013}.
Upon further doping, before the singlet state appears, the system enters another triplet $p$-wave regime in which $P^{\uparrow}_{p_x}$ and $P^{\downarrow}_{p_y}$ remain enhanced but $P^{T}_{p_y}$ is almost completely suppressed.

At intermediate doping, the $d+id$ singlet correlation $P^{S}_{d+id}$ becomes dominant while the triplet correlations are strongly suppressed. This $d+id$ window is considerably narrower than the broad intermediate-doping region expected from mean-field theory~\cite{Okamoto_PRB2013}.
More specifically, in the SU(2) slave-boson mean-field phase diagram, the antiferromagnetic Kitaev limit ($J_K=2$, $J_H=0$ in that convention) shows only a small triplet superconducting ($p{\rm SC}_1$) region up to $\delta\simeq 0.05$, and the $d+id$ state remains stable over the remaining doping range.
In our results, at higher doping, weak triplet $p$-wave correlations reappear.
However, their amplitude is small and their size dependence is large, so we cannot conclude that robust superconductivity survives in this region.

Figure~\ref{fig:sc_mall_af}(b) shows the doping dependence of the total magnetization $M_{\rm all}$.
In the two low-doping triplet regimes, $M_{\rm all}$ takes a finite value, reflecting the spin-rotational symmetry breaking in the triplet pairing.
In contrast, $M_{\rm all}$ vanishes in the intermediate-doping $d+id$ singlet regime, consistent with the spin-rotational invariance of the singlet pairing.
The antiferromagnetic triplet regimes therefore also show coexistence of triplet superconductivity and ferromagnetism, although the ferromagnetic moment is much smaller than that in the ferromagnetic case.
Toward higher doping, $M_{\rm all}$ rises strongly and the system eventually approaches an itinerant ferromagnetic state, while the superconducting correlations remain weak and size dependent.
These observations show that the pairing symmetry is controlled not only by the sign of $K$ but also by how doping reshapes the competing magnetic correlations.
For both signs of $K$, the triplet pairing always coexists with ferromagnetism, whereas the singlet $d+id$ pairing does not.

\begin{figure}[t]
  \includegraphics[width=\columnwidth]{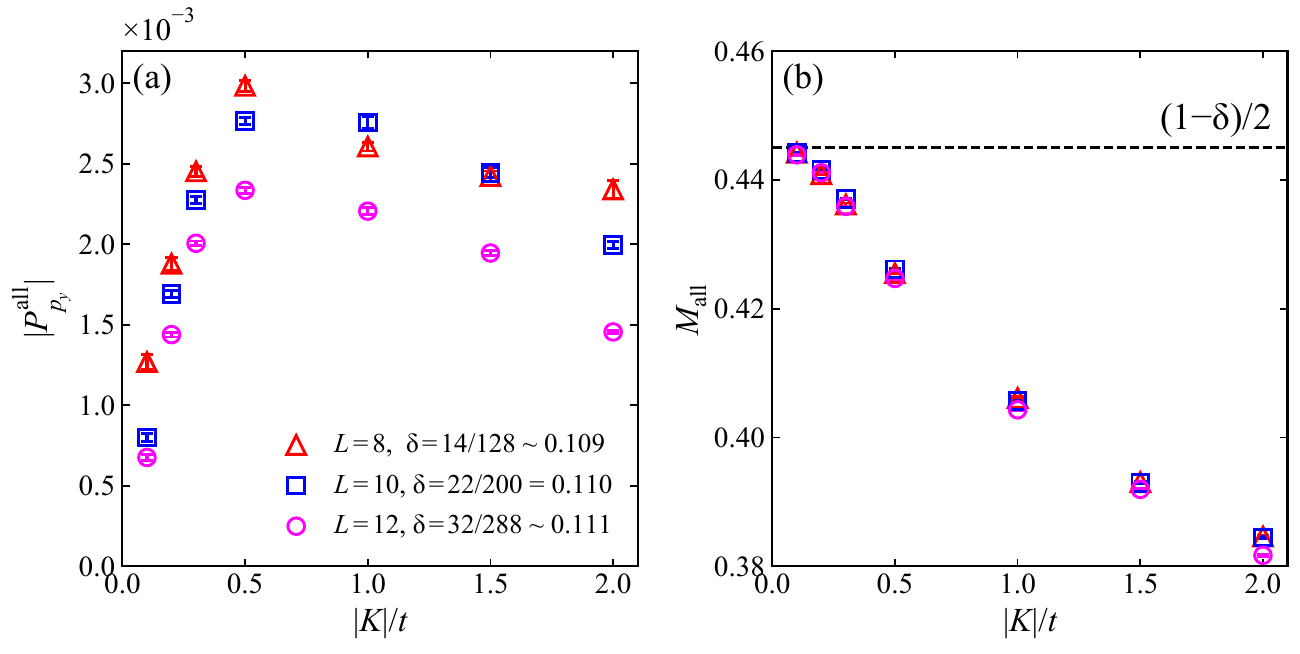}
  \caption{%
    Kitaev-coupling dependence at fixed low hole doping for the ferromagnetic interaction ($K<0$).
    (a)~Long-range-averaged triplet correlation $P^{\rm all}_{p_y}$ and
    (b)~total magnetization $M_{\rm all}$ as functions of $|K|/t$ for $L=8$, $10$, and $12$ at hole concentration $\delta\simeq 0.11$.
    The dashed line in (b) shows the fully polarized value $(1-\delta)/2$ at $\delta=0.11$.
  }
  \label{fig:kdep_ferro}
\end{figure}

Finally, we examine how triplet superconductivity
and ferromagnetism depend on the strength of
the Kitaev coupling.
We focus on the ferromagnetic interaction 
at fixed low doping $\delta\simeq 0.11$ and vary the 
coupling in the range $0.1\le |K|/t\le 2$.
Figure~\ref{fig:kdep_ferro}(a) shows the 
coupling dependence of $P^{\rm all}_{p_y}$.
The triplet correlation depends nonmonotonically 
on $|K|/t$ and has a maximum around
$|K|/t\simeq 0.5$.
In contrast, the magnetization $M_{\rm all}$ decreases 
smoothly and monotonically with $|K|/t$ and 
shows no anomaly in the same coupling 
range [Fig.~\ref{fig:kdep_ferro}(b)].

These results demonstrate that stronger Kitaev interaction does not
necessarily lead to stronger triplet superconductivity.
At the same time, ferromagnetism does not simply suppress it.
As the coupling decreases from the strong-coupling side, the triplet correlation increases together with the magnetization, and it is suppressed only when the magnetization approaches the fully polarized value in the weak-coupling limit.
The suppression of the triplet correlation near full polarization is consistent with the doping dependence at fixed coupling [Fig.~\ref{fig:sc_mall_ferro}].
The maximum at intermediate coupling thus indicates that the triplet superconductivity is optimized slightly below full polarization, where the ferromagnetic moment remains large while the spin fluctuations mediate the pairing.

{\it Summary and discussion.---}%
\label{sec:summary}
Using the many-variable variational Monte Carlo method with a projected pair-product wave function that exactly represents the Kitaev quantum spin liquid at half filling, we have elucidated the phase diagrams of the hole-doped $t$-$J$-type Kitaev model summarized in Fig.~\ref{fig:model_pr}(a).
For the ferromagnetic interaction ($K<0$), triplet $p$-wave superconductivity and ferromagnetism develop together upon doping and persist until the system approaches full polarization.
For the antiferromagnetic interaction ($K>0$), the dominant pairing changes from spin-dependent triplet $p$-wave to singlet $d+id$ as doping increases, and only weak, strongly size-dependent triplet correlations remain at high doping.
The spin- and form-factor-selective triplet pairing at low doping was not identified in the previous mean-field studies~\cite{You_PRB2012,Hyart_PRB2012,Okamoto_PRB2013}.
At fixed low doping in the ferromagnetic case, we have further shown that the triplet correlation is not simply enhanced by increasing the strength of the Kitaev coupling but has a maximum at intermediate coupling.
These results indicate that, except in the $d+id$ phase, doping simultaneously induces ferromagnetism and superconductivity irrespective of the sign of the Kitaev interaction.
The doped Kitaev model thus provides a microscopic realization of the coexistence of spin-triplet superconductivity and ferromagnetism.
Such coexistence has so far been explored mainly in $f$-electron systems~\cite{Aoki_JPSJ2019}, and doped Kitaev materials offer a different route to it.

Future studies should test the robustness of these phases by including additional interactions, such as Heisenberg and off-diagonal $\Gamma$ terms, and by allowing competing magnetic and charge orders in enlarged-unit-cell wave functions.
Evaluating the topological character of the superconducting states is left for future study.
In addition, extending the present mVMC analysis to multi-orbital Hubbard models derived from \textit{ab initio} calculations is an important challenge for exploring unconventional superconductivity in doped Kitaev candidate materials.

The authors thank Kiyu Fukui, Joji Nasu, Yasuyuki Kato, Tsuyoshi Okubo, and Yukitoshi Motome for fruitful discussions.
This work was financially supported by Grants-in-Aid for Scientific Research (KAKENHI)
(Grant Nos. JP23H03818, JP23K13055, and JP26K00652).
T.M. was supported by JST FOREST (Grant No.~JPMJFR236N).
The computations were performed using the facilities at the Supercomputer Center, Institute for Solid State Physics, University of Tokyo.

\providecommand{\noopsort}[1]{}\providecommand{\singleletter}[1]{#1}%

\clearpage
\renewcommand{\green}[1]{#1}

\onecolumngrid
\begin{center}
{\large\bfseries
Supplemental Material for ``Exotic superconductivity in the doped Kitaev quantum spin liquid''
}

\vspace{0.8em}
Takahiro Misawa$^1$ and Kota Ido$^2$

\vspace{0.4em}
{\itshape
$^1$Institute for Solid State Physics, University of Tokyo, Kashiwa, Chiba 277-8581, Japan\\
$^2$Department of Applied Science, Graduate School of Sciences and Technology for Innovation, Yamaguchi University,
2-16-1 Tokiwadai, Ube, Yamaguchi 755-8611, Japan
}
\end{center}
\twocolumngrid

\appendix
\setcounter{secnumdepth}{2}
\setcounter{figure}{0}
\renewcommand{\thefigure}{S\arabic{figure}}
\setcounter{table}{0}
\renewcommand{\thetable}{S\Roman{table}}
\section{Kitaev quantum spin liquid as a pair-product (Pfaffian) state}
\label{app:kqsl_pfaffian}

In this section, based on Refs.~\cite{SM:Burnell_Nayak_PRB2011,SM:Schaffer_2012,SM:Fu_PRB2018,SM:Udagawa_JPCM2021}, we explain how the Kitaev quantum spin liquid can be described by the projected pair-product wave function.

\subsection{Mapping between Abrikosov and Majorana fermions}
\label{appsubsec:majorana_spinon}

We start from the Kitaev model on the honeycomb lattice~\cite{SM:Kitaev_ANP2006},
\begin{align}
  \Ha_{\rm K}
  = \sum_{\langle i,j\rangle_{\alpha}} K_{\alpha} S_i^{\alpha} S_j^{\alpha},
  \label{eq:kitaev_model_sm}
\end{align}
where $S_i^{\alpha}$ is the $\alpha$ component of the spin-$1/2$ operator at site $i$
and $\langle i,j\rangle_{\alpha}$ denotes a nearest-neighbor bond of type $\alpha\in\{x,y,z\}$.
To connect the Kitaev quantum spin liquid to a projected pair-product wave function, we represent the spin operators by Abrikosov fermions (spinons), $c_{j\uparrow}$ and $c_{j\downarrow}$,
\begin{align}
  S_j^+ = c^\dagger_{j\uparrow}c_{j\downarrow},\quad
  S_j^- = c^\dagger_{j\downarrow}c_{j\uparrow},\quad
  S_j^z = \frac{1}{2}(n_{j\uparrow}-n_{j\downarrow}),
  \label{eq:abrikosov_spin}
\end{align}
together with the local single-occupancy constraint
\begin{align}
  n_{j\uparrow}+n_{j\downarrow}=1,
  \label{eq:single_occupancy}
\end{align}
which excludes the empty and doubly occupied states.

We decompose the Abrikosov fermions into four Majorana fermions $\{a_j,b_j^x,b_j^y,b_j^z\}$ at each site using an orthogonal matrix $T$,
\begin{align}
  &\begin{pmatrix}
    b^z_j \\ a_j \\ b^x_j \\ b^y_j
  \end{pmatrix}
  = T\,\mathcal{M}
  \begin{pmatrix}
    c_{j\uparrow} \\ c^\dagger_{j\uparrow} \\ c_{j\downarrow} \\ c^\dagger_{j\downarrow}
  \end{pmatrix},
  \qquad
  T^{\mathsf T}T=I,
  \label{eq:majorana_orthogonal}\\
  &\mathcal{M}=
  \begin{pmatrix}
    1 & 1 & 0 & 0 \\
    i & -i & 0 & 0 \\
    0 & 0 & 1 & 1 \\
    0 & 0 & -i & i
  \end{pmatrix},
  \label{eq:U_quadrature}
\end{align}
where $\mathcal{M}$ maps the Abrikosov fermions to the Majorana quadratures, and any orthogonal $T$ preserves the Majorana anticommutation relations.
Because the Abrikosov-fermion form \eqref{eq:abrikosov_spin} is fixed, different choices of $T$ produce apparently different Majorana representations of the same physical states.
The choice $T=I$ defines the \textit{simple} representation, for which the inverse transformation reads
\begin{align}
  \begin{pmatrix}
    c_{j\uparrow} \\
    c^\dagger_{j\uparrow} \\
    c_{j\downarrow} \\
    c^\dagger_{j\downarrow}
  \end{pmatrix}
  =\frac{1}{2}
  \begin{pmatrix}
    1 & -i & 0 & 0 \\
    1 & i & 0 & 0 \\
    0 & 0 & 1 & i \\
    0 & 0 & 1 & -i
  \end{pmatrix}
  \begin{pmatrix}
    b^z_j \\
    a_j \\
    b^x_j \\
    b^y_j
  \end{pmatrix}.
  \label{eq:spinon_from_majorana_simple}
\end{align}
Substituting Eq.~\eqref{eq:spinon_from_majorana_simple} into Eqs.~\eqref{eq:abrikosov_spin} and \eqref{eq:single_occupancy} gives Kitaev's Majorana representation,
\begin{align}
  S_j^\alpha=\frac{1+D_j}{2}\tilde{S}_j^\alpha,\ \tilde{S}_j^\alpha = \frac{i}{2} a_j b_j^\alpha,
  \label{eq:kitaev_majorana_rep}
\end{align}
and the physical constraint,
\begin{align}
  D_j = a_j b_j^x b_j^y b_j^z = 1,
  \label{eq:physical_constraint}
\end{align}
on the constrained subspace.
In terms of the Abrikosov fermions, the operator $D_j$ reduces to $D_j=-(1-2n_{j\uparrow})(1-2n_{j\downarrow})$, which takes the value $+1$ for singly occupied states and $-1$ for empty and doubly occupied states.
The constraint $D_j=1$ is therefore equivalent to the single-occupancy constraint~\eqref{eq:single_occupancy}.

Another convenient choice is the \textit{symmetric} representation~\cite{SM:Udagawa_JPCM2021},
\begin{align}
  T_{\rm sym}=\frac{1}{2}
  \begin{pmatrix}
    -1 & -1 &  1 & -1 \\
     1 & -1 &  1 &  1 \\
    -1 & -1 & -1 &  1 \\
     1 & -1 & -1 & -1
  \end{pmatrix}.
  \label{eq:T_sym}
\end{align}
By introducing $\zeta=1+i$, we write the inverse transformation as
\begin{align}
  \begin{pmatrix}
    c_{j\uparrow} \\
    c^\dagger_{j\uparrow} \\
    c_{j\downarrow} \\
    c^\dagger_{j\downarrow}
  \end{pmatrix}
  =\frac{1}{4}
  \begin{pmatrix}
    -\zeta^\ast & \zeta & -\zeta^\ast & \zeta \\
    -\zeta & \zeta^\ast & -\zeta & \zeta^\ast \\
    \zeta^\ast & \zeta & -\zeta^\ast & -\zeta \\
    \zeta & \zeta^\ast & -\zeta & -\zeta^\ast
  \end{pmatrix}
  \begin{pmatrix}
    b^z_j \\
    a_j \\
    b^x_j \\
    b^y_j
  \end{pmatrix}.
  \label{eq:spinon_from_majorana_sym}
\end{align}
This inverse transformation makes explicit that each spinon operator contains equal-weight contributions from all four Majorana fermions.
Substitution into Eqs.~\eqref{eq:abrikosov_spin} and \eqref{eq:single_occupancy} yields the same physical spin operators and single-occupancy subspace in the symmetric Majorana basis.
The Majorana operators of Ref.~[\onlinecite{SM:Udagawa_JPCM2021}] coincide with
those defined by Eq.~\eqref{eq:T_sym} up to the sign change
$(b^z,b^x,b^y)\to(-b^z,-b^x,-b^y)$.
Below, we show how the simple and symmetric Majorana representations yield two pair-product representations of the Kitaev quantum spin liquid.

\subsection{Mapping onto the Bogoliubov--de Gennes (BdG) Hamiltonian}
\label{appsubsec:bdg}

In the extended Hilbert space, the Kitaev Hamiltonian can be written as a quadratic Hamiltonian for one Majorana species
coupled to static $Z_2$ gauge fields $u_{ij}^\alpha = i b_i^\alpha b_j^\alpha=\pm 1$ on each $\alpha$ bond, where the bond $\langle i,j\rangle_\alpha$ is oriented from the $A$ to the $B$ sublattice.
Substituting Eq.~\eqref{eq:kitaev_majorana_rep}, each bond term reduces to
\begin{align}
  K_\alpha \tilde{S}_i^\alpha \tilde{S}_j^\alpha=-\frac{iK_\alpha}{4}\,a_i a_j\,u^\alpha_{ij}.
  \label{eq:kitaev_bond_majorana}
\end{align}

Because the gauge fields are static, fixing a gauge configuration maps the model onto a BdG Hamiltonian for spinons of the form shown in Eq.~\eqref{eq:kitaev_bdg_up}.
In the uniform (flux-free) sector $u_{ij}^\alpha=1$, using the simple representation [$T=I$ in Eq.~\eqref{eq:majorana_orthogonal}], we obtain a spin-polarized pairing problem for the up-spin sector,
\begin{align}
  \tilde{\Ha}_{\rm simple}(u_{ij}^\alpha=1)
  &= \frac{1}{4}\sum_{\langle i,j\rangle_\alpha}
  \notag\\[-2pt]
  &\quad\times
  \left(
    t_\alpha\, c^\dagger_{i\uparrow}c_{j\uparrow}
    +\Delta_\alpha\, c^\dagger_{i\uparrow}c^\dagger_{j\uparrow}
    +{\rm H.c.}
  \right),
  \label{eq:kitaev_bdg_up}
\end{align}
with $t_\alpha=-iK_\alpha$ and $\Delta_\alpha=iK_\alpha$.
In the symmetric representation [$T=T_{\rm sym}$ in Eq.~\eqref{eq:majorana_orthogonal}], the same flux-free Hamiltonian is expressed as
\begin{align}
  \tilde{\Ha}_{\rm sym}(u_{ij}^\alpha=1)
  &= \frac{1}{8}\sum_{\langle i,j\rangle_\alpha}\sum_{s,s'}
  \notag\\[-2pt]
  &\quad\times
  \left(
    -iK_\alpha\, c^\dagger_{is}c_{js'}
    +K_\alpha\, c^\dagger_{is}c^\dagger_{js'}
    +{\rm H.c.}
  \right),
  \label{eq:kitaev_bdg_sym}
\end{align}
in which the hopping and pairing amplitudes are distributed with equal weight over all four spin pairs.

In this study, rather than fixing a gauge sector explicitly, we use a penalty term that energetically favors
$u_{ij}^\alpha=1$ on each bond,
\begin{align}
  \tilde{\Ha}^{\alpha}_{\rm penalty}(\lambda)
  = \lambda\sum_{\langle i,j\rangle_\alpha}\left|u_{ij}^\alpha-1\right|^2
  = \lambda\sum_{\langle i,j\rangle_\alpha}\left(2-2 i b_i^\alpha b_j^\alpha\right),
  \label{eq:penalty}
\end{align}
and diagonalize the resulting quadratic Hamiltonian
\begin{align}
  \tilde{\Ha}(\lambda)
  = \tilde{\Ha}_{\rm simple/sym}(u_{ij}^\alpha=1)
  + \sum_{\alpha}\tilde{\Ha}^{\alpha}_{\rm penalty}(\lambda),
  \label{eq:quadratic_ham}
\end{align}
through the BdG equation.
Writing the Nambu spinor as $\Psi=(c_1,\ldots,c_N,c_1^\dagger,\ldots,c_N^\dagger)^{\mathsf{T}}$ (with a combined index $J\!=\!(j,\sigma)$ and $N$ denoting the number of single-particle spin orbitals),
the BdG eigenproblem takes the standard form
\begin{align}
  \begin{pmatrix}
    H & \Delta \\
    \Delta^\dagger & -H^{\mathsf{T}}
  \end{pmatrix}
  \begin{pmatrix}
    \vec{u}_n \\ \vec{v}_n
  \end{pmatrix}
  =E_n
  \begin{pmatrix}
    \vec{u}_n \\ \vec{v}_n
  \end{pmatrix},
  \label{eq:bdg}
\end{align}
where $H$ and $\Delta$ are the hopping and pairing matrices of
$\tilde{\Ha}(\lambda)$, and $\vec{u}_n$ and $\vec{v}_n$ are the particle and
hole components of the $n$-th eigenvector.

\subsection{Pair-product (Pfaffian) wave function from BdG}
\label{appsubsec:pp_pfaff}

The BdG ground state can be written as a generalized BCS state~\cite{SM:Becca_Sorella_Book2017},
\begin{align}
  \ket{\psi_{\rm BCS}(F)} =
  \exp\!\left(
    \sum_{I,J} F_{IJ}\, c_I^\dagger c_J^\dagger
  \right)\ket{0},
  \qquad F=-F^{\mathsf{T}}.
  \label{eq:bcs_state}
\end{align}
For a Bogoliubov transformation written as $\gamma = V^{\mathsf{T}} c + U^{\mathsf{T}} c^\dagger$,
the pairing matrix is obtained as
\begin{align}
  F=\frac{1}{2}UV^{-1}.
  \label{eq:F_from_UV}
\end{align}
Because the pair operators satisfy $c_I^\dagger c_J^\dagger=-c_J^\dagger c_I^\dagger$,
only the skew-symmetric part of $F$ contributes to the state,
and we take $F$ to be skew-symmetric without loss of generality.

By projecting the BCS state \eqref{eq:bcs_state} onto the sector with a fixed particle number $N_{\rm e}$,
we obtain the pair-product wave function,
\begin{align}
  \ket{\phi_{\rm pair}(F)}=
  \left(
    \sum_{I,J} F_{IJ}\, c_I^\dagger c_J^\dagger
  \right)^{N_{\rm e}/2}\ket{0},
  \label{eq:pp_state}
\end{align}
which is the form used in mVMC.
In real-space Monte Carlo sampling, the amplitude $\braket{x|\phi_{\rm pair}(F)}$ for a configuration $\ket{x}$
is evaluated as a Pfaffian of a submatrix of $F$, which is why the pair-product wave function is also referred to as the Pfaffian wave function.

\subsection{Kitaev quantum spin liquid in mVMC}
\label{appsubsec:kitaevqsl_mvmc}

The variational wave function for the Kitaev quantum spin liquid (KQSL) is constructed by projecting $\ket{\phi_{\rm pair}(F)}$ onto the physical Hilbert space and is used as the starting point for the doped problem.
At half filling (spin model), we impose the single-occupancy constraint \eqref{eq:single_occupancy},
\begin{align}
  \ket{\Phi_{\rm KQSL}}
  =\PG^{\infty}\,\ket{\phi_{\rm pair}(F_{\rm KQSL})},
  \label{eq:kqsl_projected}
\end{align}
where $F_{\rm KQSL}$ can be chosen from the BdG solution of $\tilde{\Ha}(\lambda)$.
For the doped $t$-$J$-type Kitaev model, we instead use the no-double-occupancy projector appropriate for the $t$-$J$ model and fix $N_{\rm e}$.
In mVMC, the matrix elements $F_{IJ}$ are treated as variational parameters and optimized.

\begin{figure}[t]
  \includegraphics[width=\columnwidth]{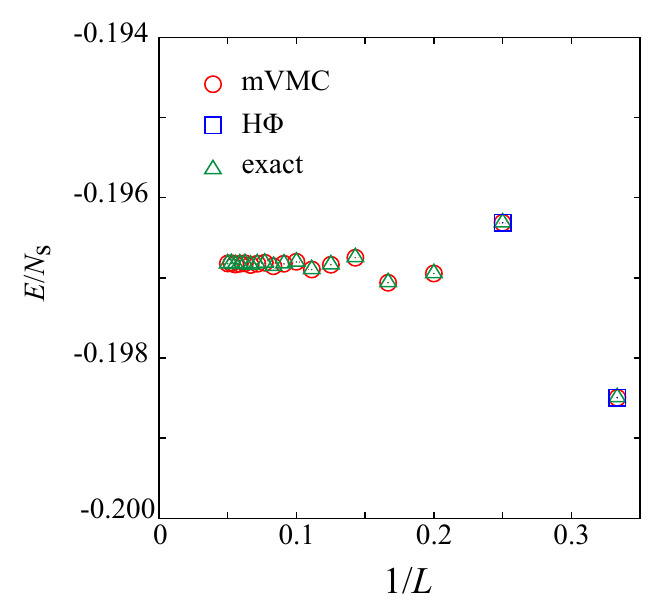}
  \caption{Ground-state energy per site $E/\Ns$ of the honeycomb Kitaev model at half filling ($K/t=-1$) versus $1/L$
  ($\Ns=2L^2$).
  Open circles: mVMC pair-product wave function with statistical errors (smaller than the symbol size).
  Squares: $\mathcal{H}\Phi$ exact diagonalization ($L=3,4$).
  Triangles: Majorana exact diagonalization.}
  \label{fig:half_filling_E}
\end{figure}

To validate that the projected pair-product wave function $\ket{\Phi_{\rm KQSL}}$ can exactly represent the Kitaev ground state,
we benchmark its variational energy at half filling against exact energies obtained by Majorana-fermion diagonalization
and, for small clusters, by full exact diagonalization with $\mathcal{H}\Phi$~\cite{SM:HPhi_v1,SM:HPhi_v2}.
Figure~\ref{fig:half_filling_E} shows that the mVMC energy reproduces the exact value up to $L=20$ within the combined Monte Carlo and optimization uncertainties, which are of order $10^{-6}$ per site.
The mVMC calculations use AP--P boundary conditions, implemented by taking $u_{ij}^{\alpha}=+1$ in the bulk and reversing its sign on bonds crossing the antiperiodic boundary. For every isotropic $L\times L$ cluster with $L\geq 3$ examined here, the AP--P sector yields the lowest Majorana energy consistent with the fermion-parity constraint, although the number of degenerate ground-state sectors depends on $L$.

\section{Superconducting ans\"atze for initial states}
\label{app:initial_states}

\subsection{Definitions of the initial superconducting states}
\label{appsubsec:initial_states_def}

In addition to the KQSL initial states constructed from
the simple and symmetric representations, we employ two superconducting
states proposed in mean-field studies of the doped Kitaev
model~\cite{SM:You_PRB2012,SM:Hyart_PRB2012,SM:Okamoto_PRB2013,SM:Scherer_PRB2014},
namely the singlet $d+id$ state and the triplet $p{\rm SC}_2$ state.
These superconducting states are obtained by diagonalizing BdG Hamiltonians
with nearest-neighbor hopping and the following pairing amplitudes.
The gap structure of the $d+id$ state is defined by the singlet pairing
amplitudes on the $(x,y,z)$ bonds,
\begin{align}
  (\Delta^{S}_{x},\,\Delta^{S}_{y},\,\Delta^{S}_{z})
  =\Delta_{0}\,(\omega,\,\omega^{\ast},\,1),
  \label{eq:did_seed}
\end{align}
with $\omega=e^{2\pi i/3}$, corresponding to the $d+id$ form factor
[Fig.~\ref{fig:model_pr}(b)].
Following Ref.~[\onlinecite{SM:Okamoto_PRB2013}], the $p{\rm SC}_2$ state is defined by the $d$-vector on
each bond $\rho=x,y,z$ through
\begin{align}
  \Delta^{\uparrow}_{\rho}&=-d^{x}_{\rho}+i d^{y}_{\rho},
  \nonumber\\
  \Delta^{\downarrow}_{\rho}&=d^{x}_{\rho}+i d^{y}_{\rho},
  \nonumber\\
  \Delta^{T}_{\rho}&=d^{z}_{\rho},
  \label{eq:psc2_gap}
\end{align}
where $d^{x}_{\rho}$ denotes the $x$ component of $\vec{d}_{\rho}$.
The $d$-vector on each $\rho$ bond has a longitudinal component
$t_1$ along the spin axis $\rho$ and transverse components $\pm t_2$.
The $d$-vectors for the ferromagnetic Kitaev interaction ($K<0$) are
defined by
\begin{align}
  \vec d_x&=(t_1,t_2,-t_2),
  \nonumber\\
  \vec d_y&=(-t_2,t_1,t_2),
  \nonumber\\
  \vec d_z&=(t_2,-t_2,t_1).
  \label{eq:psc2_d_fm}
\end{align}
For the antiferromagnetic Kitaev interaction ($K>0$), the $d$-vectors
are given by
\begin{align}
  \vec d_x&=(t_1,t_2,t_2),
  \nonumber\\
  \vec d_y&=(t_2,t_1,t_2),
  \nonumber\\
  \vec d_z&=(t_2,t_2,t_1).
  \label{eq:psc2_d_afm}
\end{align}
We fix $\Delta_0=t_1=5$ and vary $t_2=-4,-3,-2,-1,0,1,2,3,4$,
which covers a wide parameter range including the $p{\rm SC}_2$ states
obtained in Ref.~[\onlinecite{SM:Okamoto_PRB2013}].

We diagonalize these BdG Hamiltonians and
obtain the pair matrix $F_{IJ}$ through Eq.~\eqref{eq:F_from_UV}.
We add small random perturbations to $F_{IJ}$ and optimize the pair-product wave
function by the stochastic reconfiguration method.
To reduce sensitivity to local minima, we also initialize the optimization with variational parameters previously optimized at nearby or representative hole concentrations. For each system size and hole concentration, we verify convergence and retain the lowest-energy state among the resulting candidates.

We summarize the initial states used to obtain the results in the
main text.
For the ferromagnetic Kitaev interaction, the triplet superconducting states
in the low-doping regime are mainly derived from the symmetric representation
of the KQSL, and the weakly magnetized superconducting states around
$\delta\simeq 0.12$ and $0.2$ are derived from the $p{\rm SC}_2$ state.
For the antiferromagnetic Kitaev interaction, the states in the low-doping
regime are derived from the simple representation of the KQSL.
The spin-dependent triplet pairing in this regime, whose form factors differ
between the up- and down-spin sectors, reflects the gap structure of the
simple representation shown in Table~\ref{tab:seed_closed_form}.

\subsection{Symmetry of the gap structures}
\label{appsubsec:seed_closed_form}

\begin{table*}[t]
\caption{Symmetry decomposition of the bare pairing amplitudes
$\Delta$ used for the KQSL and $p{\rm SC}_2$ initial states.
The entries are $|A^\eta_\alpha|^2$ in the nearest-neighbor pairing channels.
For the simple and symmetric representations, $K$ and $\lambda$ denote
the Kitaev and penalty coefficients in the BdG Hamiltonian.
For the $S_z=0$ triplet entries, we define the bond amplitude as
$\Delta^T_\rho\equiv\Delta_{\rho,\uparrow\downarrow}
=\Delta_{\rho,\downarrow\uparrow}=d_\rho^z$, without the $\sqrt{2}$
normalization used for the pair operator in Eq.~\eqref{eq:pair_op_ST}.}
\label{tab:seed_closed_form}
\begin{ruledtabular}
\begin{tabular}{llcccc}
Initial state & Spin channel $\eta$ & $f$-wave & $p_x$ & $p_y$ & $p+ip$ \\
\hline
simple & $\uparrow\uparrow$ &
$(3K/4+2\lambda)^2$ & $0$ & $16\lambda^2$ & $4\lambda^2$ \\
simple & $\downarrow\downarrow$ &
$0$ & $16\lambda^2$ & $0$ & $12\lambda^2$ \\
simple & $T$ ($S_z=0$) &
$0$ & $0$ & $0$ & $0$ \\
\hline
symmetric & $\uparrow\uparrow,\downarrow\downarrow$ &
$(3K/8+\lambda)^2$ & $4\lambda^2$ & $4\lambda^2$ & $4\lambda^2$ \\
symmetric & $T$ ($S_z=0$) &
$(3K/8+\lambda)^2$ & $0$ & $16\lambda^2$ & $4\lambda^2$ \\
\hline
$p{\rm SC}_2$ ($K>0$) & $T$ ($S_z=0$) &
$(t_1+2t_2)^2$ & $0$ & $4(t_1-t_2)^2$ & $(t_1-t_2)^2$ \\
$p{\rm SC}_2$ ($K>0$) & $\uparrow\uparrow$ &
$2(t_1+2t_2)^2$ & $2(t_1-t_2)^2$ & $2(t_1-t_2)^2$ & $(2+\sqrt{3})(t_1-t_2)^2$ \\
$p{\rm SC}_2$ ($K>0$) & $\downarrow\downarrow$ &
$2(t_1+2t_2)^2$ & $2(t_1-t_2)^2$ & $2(t_1-t_2)^2$ & $(2-\sqrt{3})(t_1-t_2)^2$ \\
\hline
$p{\rm SC}_2$ ($K<0$) & $T$ ($S_z=0$) &
$t_1^2$ & $4t_2^2$ & $4t_1^2$ & $t_1^2+3t_2^2$ \\
$p{\rm SC}_2$ ($K<0$) & $\uparrow\uparrow$ &
$2t_1^2$ & $2(t_1^2+t_2^2)$ & $2(t_1^2+9t_2^2)$ & $(2+\sqrt{3})(t_1^2+3t_2^2)$ \\
$p{\rm SC}_2$ ($K<0$) & $\downarrow\downarrow$ &
$2t_1^2$ & $2(t_1^2+t_2^2)$ & $2(t_1^2+9t_2^2)$ & $(2-\sqrt{3})(t_1^2+3t_2^2)$ \\
\end{tabular}
\end{ruledtabular}
\end{table*}

Table~\ref{tab:seed_closed_form} summarizes the bare nearest-neighbor
pairing amplitudes of the simple, symmetric, and $p{\rm SC}_2$ initial states.
We define the channel decomposition
\begin{align}
  A^{\eta}_{\alpha}=\sum_{\rho} f^{\alpha}_{\rho}\Delta^{\eta}_{\rho},
  \label{eq:channel_decomposition}
\end{align}
where $\rho$ runs over the three nearest-neighbor bond types $\{x,y,z\}$ and
$\vec{f}^{\,\alpha}=(f^{\alpha}_{x},f^{\alpha}_{y},f^{\alpha}_{z})$ is the form factor
for channel $\alpha$ [Eq.~\eqref{eq:pair_op_ST}].
The form factors are given by
$\vec{f}^{\,f}=(1,1,1)$,
$\vec{f}^{\,p_x}=(1,-1,0)$,
$\vec{f}^{\,p_y}=(1,1,-2)$, and
$\vec{f}^{\,p+ip}=(\omega,\omega^{\ast},1)$ with $\omega=e^{2\pi i/3}$.

The variational calculation is initialized by the pairing matrix $F$ obtained
from the BdG ground state through Eq.~\eqref{eq:F_from_UV}.
Because $F=\frac{1}{2}UV^{-1}$ is determined by the full BdG Hamiltonian
rather than by $\Delta$ alone, its channel decomposition generally differs
from that of the bare pairing matrix $\Delta$.
For example, in the simple representation, the $\downarrow\downarrow$ component of $\Delta$
is a pure $p_x$ combination, whereas the corresponding component of $F$ can have
equal $f$-wave and $p_y$ weights and no $p_x$ weight.
The entries in Table~\ref{tab:seed_closed_form} therefore label the
construction of the initial states and do not directly represent the channel content of the initial
variational states.

\section{Details of pairing correlations}
\label{app:pairing_details}

\subsection{Representative real-space pairing correlations}
\label{appsubsec:real_space_Pr}

\begin{figure*}[t]
  \begin{minipage}[t]{0.32\textwidth}
    \includegraphics[width=\textwidth]{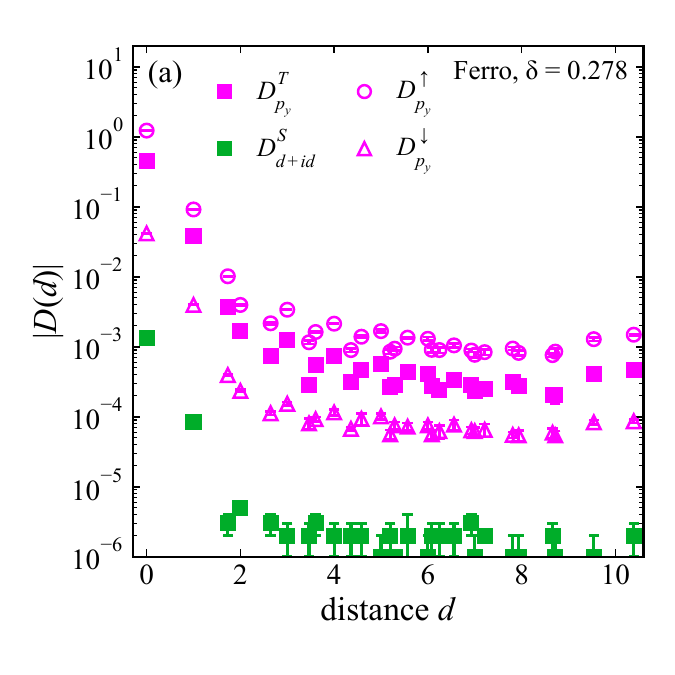}
  \end{minipage}\hfill
  \begin{minipage}[t]{0.32\textwidth}
    \includegraphics[width=\textwidth]{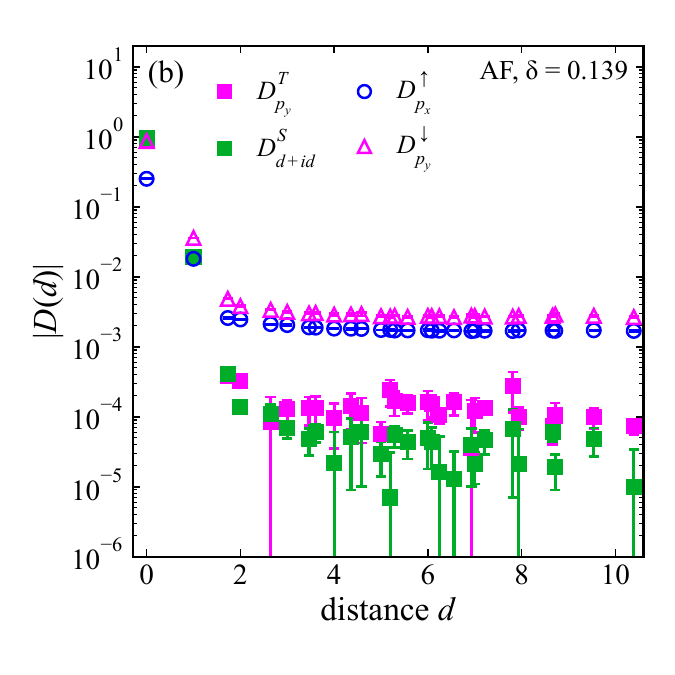}
  \end{minipage}\hfill
  \begin{minipage}[t]{0.32\textwidth}
    \includegraphics[width=\textwidth]{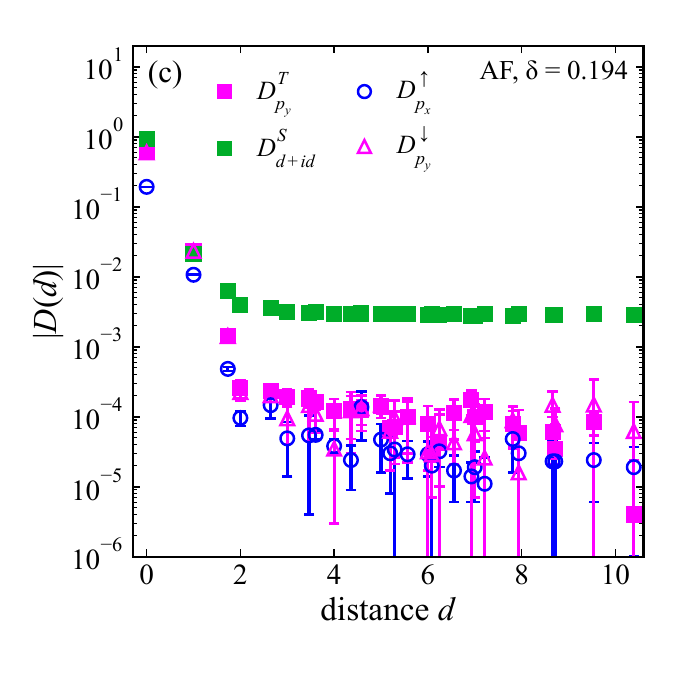}
  \end{minipage}
  \caption{Distance dependence of the real-space pairing correlations $|D^\eta_\alpha(\vec{r})|$ on the $L=12$ cluster ($\Ns=288$), shown on a logarithmic scale.
  (a)~Ferromagnetic Kitaev interaction ($K/t=-1$) at higher doping, $\delta=0.278$.
  (b)~Antiferromagnetic Kitaev interaction ($K/t=+1$) in the equal-spin triplet regime, $\delta=0.139$.
  (c)~Antiferromagnetic Kitaev interaction in the $d+id$ singlet regime, $\delta=0.194$.
  Each channel is identified by color and marker style: the form factor is encoded by color ($p_x$ blue, $p_y$ magenta), and the spin sector by marker style ($S_z=0$ filled squares, up-spin open circles, down-spin open triangles). The singlet $d+id$ channel $D^{S}_{d+id}$ is shown as green filled squares.}
  \label{fig:real_space_Pr}
\end{figure*}

For the distance-resolved pairing correlator $D^\eta_\alpha(\vec{r})$ in Eq.~\eqref{eq:pair_corr},
we parameterize the displacement as
$\vec{r}=d_x\vec{a}_1+d_y\vec{a}_2$, where
$\vec{a}_1=(1,0)$ and $\vec{a}_2=(1/2,\sqrt{3}/2)$.
After independently reducing $d_x$ and $d_y$ to $(-L/2,L/2]$, we assign
the distance $d=|\vec{r}|=(d_x^2+d_xd_y+d_y^2)^{1/2}$.
For each distinct distance $d$, we define $D^\eta_\alpha(d)$ as the value
of $\operatorname{Re}D^\eta_\alpha(\vec{r})$ with the largest absolute
value over all $\vec{r}$ satisfying $|\vec{r}|=d$.
The long-range correlation in Eq.~\eqref{eq:pair_avg} is the average of
these representative values over the distances satisfying $d\ge L/2$.
The long-range averages $P^\eta_\alpha$ in the main text can be interpreted as measures of long-range order when the underlying real-space correlators $D^\eta_\alpha(\vec{r})$ approach finite plateaus at large distances rather than decaying.
Figure~\ref{fig:real_space_Pr} shows representative distance profiles $|D^\eta_\alpha(\vec{r})|$ on the $L=12$ cluster for the three characteristic regimes identified in the main text.
For the ferromagnetic Kitaev case at higher doping [panel~(a), $\delta=0.278$], the $p_y$ triplet channels $D^{\uparrow}_{p_y}$ and $D^{T}_{p_y}$ remain long-ranged, while the singlet $d+id$ correlation drops below $10^{-5}$.
In the antiferromagnetic equal-spin triplet regime [panel~(b), $\delta=0.139$], the equal-spin channels $D^{\uparrow}_{p_x}$ and $D^{\downarrow}_{p_y}$ approach a plateau of order $10^{-3}$, while the $S_z=0$ triplet $D^{T}_{p_y}$ and the singlet $D^{S}_{d+id}$ become indistinguishable from zero within the statistical resolution.
In the antiferromagnetic intermediate-doping regime [panel~(c), $\delta=0.194$], only the singlet $d+id$ channel retains a long-range plateau, whereas all triplet channels decay.
These profiles are consistent with long-range pairing order.

\subsection{\texorpdfstring{Full $p$-wave triplet-channel decomposition}{Full p-wave triplet-channel decomposition}}
\label{appsubsec:triplet_channels}

\begin{figure*}[t]
  \begin{minipage}[t]{0.32\textwidth}
    \includegraphics[width=\textwidth]{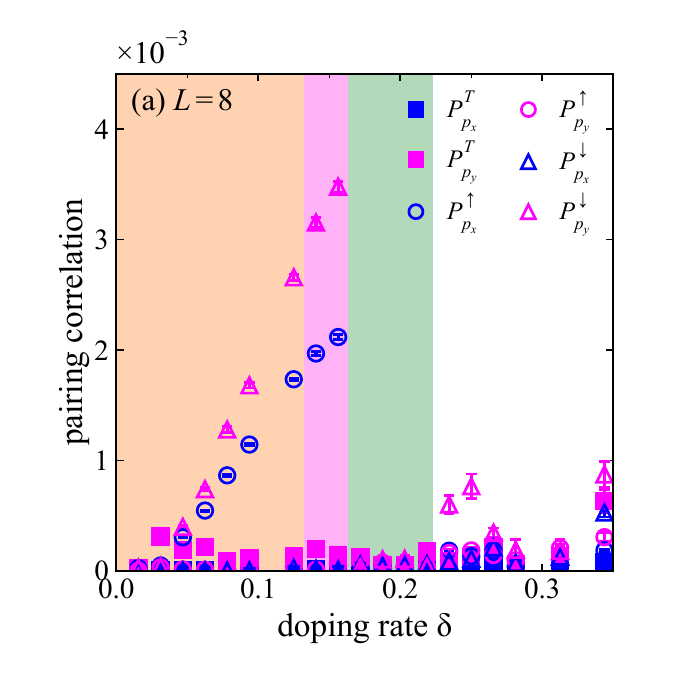}
  \end{minipage}\hfill
  \begin{minipage}[t]{0.32\textwidth}
    \includegraphics[width=\textwidth]{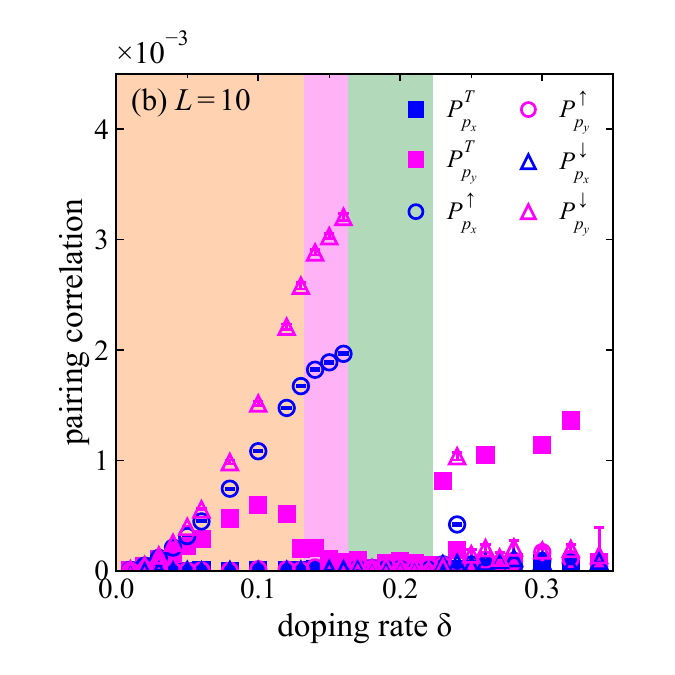}
  \end{minipage}\hfill
  \begin{minipage}[t]{0.32\textwidth}
    \includegraphics[width=\textwidth]{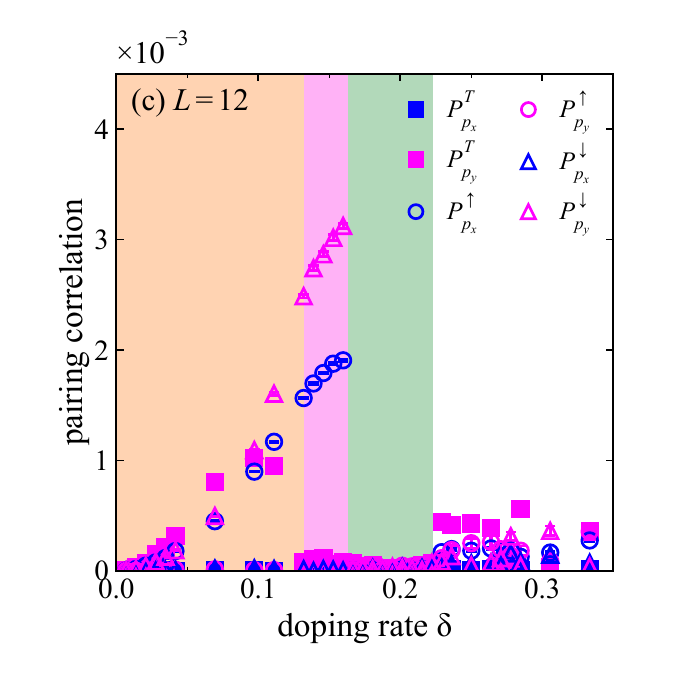}
  \end{minipage}
  \caption{Doping dependence of the six spin- and form-factor-resolved triplet pairing correlations for the antiferromagnetic Kitaev interaction ($K/t=+1$) on the (a)~$L=8$, (b)~$L=10$, and (c)~$L=12$ clusters.
  The form factor is encoded by color ($p_x$ blue, $p_y$ magenta) and the spin sector by marker style ($S_z=0$ triplet filled squares, up-spin open circles, down-spin open triangles), so that all six channels $P^{T}_{p_x}$, $P^{T}_{p_y}$, $P^{\uparrow}_{p_x}$, $P^{\uparrow}_{p_y}$, $P^{\downarrow}_{p_x}$, and $P^{\downarrow}_{p_y}$ are distinguished.
  The shaded regions indicate the same phase regions as in Fig.~\ref{fig:sc_mall_af}.}
  \label{fig:triplet_6channels}
\end{figure*}

Figure~\ref{fig:sc_mall_af}(a) in the main text shows only the representative pairing channels that become dominant in each doping regime.
Figure~\ref{fig:triplet_6channels} presents the complete decomposition of the $p_x$- and $p_y$-wave triplet correlations into all six spin- and form-factor-resolved components, namely the $S_z=0$ triplet ($P^{T}_{p_x}$, $P^{T}_{p_y}$) and the equal-spin triplet ($P^{\uparrow}_{p_x}$, $P^{\uparrow}_{p_y}$, $P^{\downarrow}_{p_x}$, $P^{\downarrow}_{p_y}$), for the antiferromagnetic Kitaev case on the $L=8$, $10$, and $12$ clusters.
The decomposition makes explicit the spin- and form-factor-selective enhancement described in the main text.
At low doping, $P^{\downarrow}_{p_y}$, $P^{\uparrow}_{p_x}$, and $P^{T}_{p_y}$ are selectively enhanced, while the corresponding channels $P^{\downarrow}_{p_x}$, $P^{\uparrow}_{p_y}$, and $P^{T}_{p_x}$ remain small.
The same selection pattern is observed for all three cluster sizes.

\clearpage

\begingroup
\makeatletter
\let\@FMN@list\@empty
\let\label\@gobble
\makeatother

\endgroup

\end{document}